\documentclass[10pt,twocolumn,letterpaper]{article}
\usepackage[pagenumbers]{wacv}
\usepackage[T1]{fontenc}
\usepackage{bm}
\usepackage[algo2e,ruled]{algorithm2e}
\usepackage{placeins}
\definecolor{wacvblue}{rgb}{0.21,0.49,0.74}
\usepackage[breaklinks,colorlinks,allcolors=wacvblue]{hyperref}
\newcommand{\ILD}{\mathbf{I}_{\mathrm{LD}}}
\newcommand{\IFD}{\mathbf{I}_{\mathrm{FD}}}
\newcommand{\IFDh}{\hat{\mathbf{I}}_{\mathrm{FD}}}
\newcommand{\Rsino}{\mathbf{R}}
\newcommand{\Rad}{\mathcal{R}_{\Theta}}
\newcommand{\Radi}{\mathcal{R}_{\Theta}^{\dagger}}
\newcommand{\Lp}[1]{\mathcal{L}_{\mathrm{#1}}}
\renewcommand{\vec}[1]{\bm{#1}}

\def\confName{WACV}
\def\confYear{2027}
\title{ProFed: Projection-Guided Personalized Federated Learning for Low-Dose CT Denoising}
\author{Anas Zafar\quad Muhammad Waqas\quad Amgad Muneer\\
Rukhmini Bandyopadhyay\quad Eman Showkatian \quad Jia Wu\\\\
The University of Texas MD Anderson Cancer Center\\
Houston, TX, USA}

\begin{document}
\maketitle

\begin{abstract}
Low-dose computed tomography (CT) reduces radiation exposure but introduces noise and artifacts whose appearance depends on acquisition conditions. Federated learning enables collaborative restoration without centrally pooling images, yet variation in anatomy and scanning protocols complicates learning a shared denoiser. We present ProFed, a personalized federated framework that combines image-domain feature modulation with projection-domain supervision. Anatomy and protocol descriptors condition a shared encoder--decoder, while differentiable forward-projection, backprojection, and cycle-consistency losses encourage agreement with client-local measurements. An uncertainty-weighted aggregation rule combines shared parameters and retains anatomy-specific heads locally. Following the SCAN-PhysFed benchmark, we train on eight Mayo Clinic patients and evaluate two held-out patients under eight known acquisition protocols. ProFed achieves protocol-averaged PSNR of 42.56~dB with RED-CNN and 44.83~dB with Uformer, exceeding SCAN-PhysFed by 1.42~dB and 1.18~dB, respectively. Per-protocol measurements and qualitative comparisons show both improvements and remaining failure cases. The results support projection-guided supervision as a promising component of federated image restoration within this controlled setting; independent deployment validation and larger multi-institutional studies remain necessary.
\end{abstract}

\section{Introduction}
\label{sec:intro}
Low-dose computed tomography (LDCT) presents a practical image-quality problem: reducing exposure also reduces measurement quality, and reconstruction can amplify noise and produce structured artifacts. Learned restoration methods can improve fidelity to full-dose references, but their behavior depends on the acquisition and image distributions encountered during training \citep{chen2017low,yang2018low,gao2023corediff}. A denoiser must balance noise suppression against preservation of low-contrast structures; an apparently smoother image is not necessarily a more faithful reconstruction.

Collaborative training is attractive because a single institution may not cover the range of anatomy and acquisition conditions needed for reliable restoration. Federated learning (FL) exchanges model updates while retaining images at participating sites \citep{mcmahan2017communication,darzidehkalani2022federated}. However, differing protocols and patient populations create statistical heterogeneity. Personalized FL addresses this by combining shared representations with client-specific adaptation \citep{arivazhagan2019federated,li2021fedbn,feng2022specificity}. For LDCT, frequency decomposition and physics-aware conditioning further tailor the model to reconstruction \citep{chen2024fedfdd,yang2025patient}.

An important design choice is the source of supervision. Image-domain losses compare the reconstructed output with a reference, whereas projection-domain losses also constrain its consistency with the acquisition. CT measurements encode attenuation along viewing rays; changes in photon statistics and angular sampling affect these measurements before reconstruction. We therefore investigate whether client-local projection supervision can complement personalized image restoration. This motivation does not require anatomy and acquisition effects to be statistically independent in a sinogram.

We introduce \textbf{ProFed}, which modulates image-domain features using anatomy and protocol descriptors and trains the resulting reconstruction with measurement-consistency losses. The distinction is deliberate: personalization acts on latent image features, while the additional supervision acts through projection operators. Each client evaluates these operators using its own acquisition geometry. Shared reconstruction and protocol parameters are aggregated, and anatomy-specific modulation and gating parameters remain local.

The paper makes three contributions:
\begin{enumerate}
\item A federated restoration formulation that combines anatomy- and protocol-conditioned feature modulation with client-local projection supervision.
\item A training objective incorporating forward-projection, backprojection, and cycle consistency, together with selective uncertainty-weighted aggregation.
\item An evaluation with RED-CNN and Uformer backbones that reports all available client-level comparisons and examines the scope of the observed image-quality gains.
\end{enumerate}
The contribution is a practical combination of acquisition-aware supervision and personalized FL, evaluated under simulated heterogeneity.

\section{Related Work}
\label{sec:related}
\paragraph{Learning for LDCT restoration.}
Residual encoder--decoder networks \citep{chen2017low}, adversarial objectives \citep{yang2018low}, transformers \citep{zhang2021transct,wang2022uformer}, and diffusion models \citep{gao2023corediff} provide different restoration priors. Self-supervised methods reduce dependence on paired targets \citep{bera2023self}. Dual-domain reconstruction and learned iterative methods incorporate acquisition information beyond an image-only objective \citep{lin2019dudonet,li2019super}. ProFed uses paired supervision and studies how measurement consistency can be incorporated into personalized federated training.

\paragraph{Federated and personalized reconstruction.}
FedAvg aggregates local updates \citep{mcmahan2017communication}; FedProx and FedNova address aspects of heterogeneous federated optimization \citep{li2020federated,wang2020tackling}. Personalization can retain local layers or normalization statistics \citep{arivazhagan2019federated,li2021fedbn}, preserve client-specific reconstruction features \citep{feng2022specificity}, or generate adaptation parameters with hypernetworks \citep{yang2023hypernetwork}. FedFDD decomposes frequency components for federated LDCT denoising \citep{chen2024fedfdd}. SCAN-PhysFed combines anatomy- and scanning-level conditioning for personalized LDCT restoration \citep{yang2025patient}. Our main distinction is the use of client-local projection losses to supervise a modulated reconstruction network.

\section{Method}
\label{sec:method}
\begin{figure*}[t]
\centering
\includegraphics[width=\textwidth]{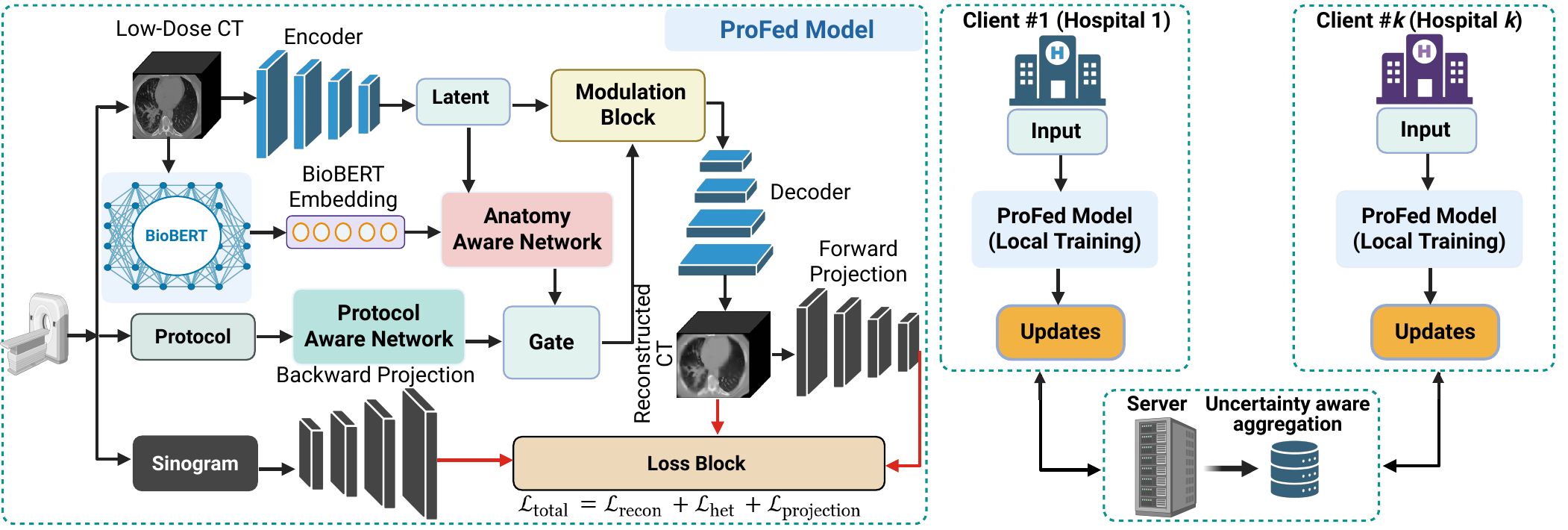}
\caption{ProFed architecture. The encoder--decoder is modulated by anatomy and protocol descriptors. Anatomy conditioning includes image features and BioBERT-encoded descriptions. Projection consistency supervises local training; the server aggregates shared parameters using client uncertainty estimates. Hospital labels illustrate the intended distributed setting; the experiments use simulated patient--protocol clients.}
\label{fig:method}
\end{figure*}

\subsection{Inputs and reconstruction model}
Client $k\in\{1,\ldots,K\}$ holds paired low- and full-dose images, a sinogram, and acquisition metadata:
\begin{equation}
\mathcal{D}_k=\{(\ILD^{(i)},\IFD^{(i)},\Rsino^{(i)},\vec p^{(i)})\}_{i=1}^{N_k}.
\end{equation}
Here $\Rsino\in\mathbb{R}^{N_\theta\times N_d}$ denotes projection measurements. Following the benchmark protocol \citep{yang2025patient}, $\vec p\in[0,1]^7$ encodes view count, detector-bin count, pixel length, detector-bin length, source-to-center distance, detector-to-center distance, and incident photon count. Each component is min--max normalized across the known protocol configurations. The reconstruction is
\begin{equation}
\IFDh=f(\ILD,\vec p,\vec a),
\end{equation}
where $\vec a\in\mathbb{R}^{768}$ is an anatomy descriptor. The sinogram provides training supervision rather than an additional input to the image encoder.

The encoder $E_\theta$ produces a spatial latent representation $\vec z$ with $C=192$ modulation channels. Protocol and anatomy branches generate channel-wise affine coefficients, and a shared decoder $D_\phi$ reconstructs the image from the modulated latent. Figure~\ref{fig:method} summarizes the architecture.

\subsection{Anatomy and protocol conditioning}
\paragraph{Anatomy descriptor.}
The anatomy pathway fuses visual features and a generated description. For a descriptor-source image $I$, a visual backbone provides $\vec a_{\mathrm{img}}\in\mathbb{R}^{512}$; an image-captioning model produces text encoded with BioBERT \citep{lee2020biobert}. The fused descriptor is
\begin{align}
\vec a_{\mathrm{txt}}&=\mathrm{BioBERT}(\mathrm{Captioner}(I)),\\
\vec a&=\mathrm{MLP}_{\mathrm{fus}}([\vec a_{\mathrm{img}};\vec a_{\mathrm{txt}}]).
\end{align}
During training, descriptors are computed from the full-dose training images and cached before federated optimization. A local anatomy head gives $(\vec\gamma_a,\vec\beta_a)=H_\alpha(\vec a)$. This pathway is retained as part of the method; its contribution is not established by a separate image-only versus image-and-text experiment.

\paragraph{Protocol descriptor.}
The protocol branch embeds the metadata as $\vec e_p=\mathrm{MLP}_\xi(\vec p)$ and predicts $(\vec\gamma_p,\vec\beta_p)=H_\beta(\vec e_p)$. A rank-8 low-rank update, $\mathbf W=\mathbf W_0+\mathbf B\mathbf A$, is used for decoder adaptation. This parameterization provides a compact adaptation mechanism; it does not by itself establish a particular end-to-end runtime or communication saving.

\paragraph{Gated modulation.}
A learned scalar gate $g=\sigma(\mathrm{FC}_g(\vec a))$ interpolates between conditioning sources:
\begin{align}
\vec\gamma&=g\vec\gamma_a+(1-g)\vec\gamma_p,\\
\vec\beta&=g\vec\beta_a+(1-g)\vec\beta_p,\\
\IFDh&=D_\phi(\vec\gamma\odot E_\theta(\ILD)+\vec\beta).
\label{eq:reconstruction}
\end{align}
The gate is a learned mixing coefficient, not a calibrated detector of incorrect descriptions. Keeping $H_\alpha$ and $\mathrm{FC}_g$ local avoids directly averaging these client-specific heads. Their influence on the shared reconstruction gradients can nevertheless propagate through aggregation.

\subsection{Projection-guided supervision}
\label{sec:projection}
We use differentiable forward projection $\Rad$ and filtered backprojection $\Radi$, implemented through coordinate rotation and bilinear interpolation. $\Theta$ denotes the client's view set. Three terms relate the output to its acquisition:
\begin{align}
\Lp{fwd}&=\|\Rad(\IFDh)-\Rsino\|_2^2,\label{eq:fwd}\\
\Lp{bwd}&=\|\IFDh-\Radi(\Rsino)\|_2^2,\label{eq:bwd}\\
\Lp{cyc}&=\|\IFDh-\Radi(\Rad(\IFDh))\|_2^2.\label{eq:cyc}
\end{align}
The projection loss is $\Lp{proj}=\lambda_f\Lp{fwd}+\lambda_b\Lp{bwd}+\lambda_c\Lp{cyc}$. The equal-weight setting uses $\lambda_f=\lambda_b=\lambda_c=1$.

Forward consistency penalizes disagreement with measured projections. Backprojection consistency anchors the estimate to an analytic reconstruction of those measurements. Since the measurements can be noisy, these terms must be balanced against full-dose image supervision; neither is a guarantee of a noise-free result. Cycle consistency penalizes change under the discrete projection--reconstruction composition. Its behavior depends on sampling, filtering, and discretization, and we do not assume that $\Radi\Rad$ is an exact inverse or an orthogonal projector.

Projection losses are evaluated locally in each client's measurement coordinates. The server combines network parameters, not sinograms or reconstructed images, so this formulation does not require cross-client alignment of projection arrays.

\subsection{Objective and federated optimization}
Image fidelity is measured by $\Lp{rec}=\|\IFDh-\IFD\|_2^2$. A variance-prediction network $\sigma^2=\mathrm{VarianceNet}(\ILD,\vec p)$ supplies a heteroscedastic term:
\begin{equation}
\Lp{het}=\sum_{x,y}\left[\frac{(\IFDh-\IFD)^2}{2\sigma^2}+\frac{1}{2}\log\sigma^2\right](x,y).
\end{equation}
The local training objective combines all three components:
\begin{equation}
\mathcal L=\lambda_r\Lp{rec}+\lambda_h\Lp{het}+\lambda_p\Lp{proj}.
\label{eq:objective}
\end{equation}

After local optimization, client $k$ estimates predictive variance using Monte Carlo dropout on its validation data:
\begin{equation}
u_k=\mathbb E_{\mathcal D_k^{\mathrm{val}}}\!\left[\operatorname{Var}_{m=1}^{M}(\IFDh^{(m)})\right].
\end{equation}
The server receives the shared weights, $u_k$, and local sample count $N_k$. Normalized aggregation weights are
\begin{equation}
w_k=\frac{N_k/(u_k+\epsilon)}{\sum_j N_j/(u_j+\epsilon)},\qquad
\theta^{\mathrm{sh}}\leftarrow\sum_k w_k\theta_k^{\mathrm{sh}},
\label{eq:aggregation}
\end{equation}
with $\epsilon>0$. Shared components comprise the encoder, decoder, and protocol-conditioning branch; the anatomy head and gate remain local. Equal uncertainty recovers sample-count weighting. Lower predictive variance increases a client's relative weight, but need not imply better calibration or smaller reconstruction bias.

Algorithm~\ref{alg:profed} summarizes the update sequence. Projection operators are absent from the reconstruction forward pass once conditioning inputs are available. End-to-end deployment cost must still include generation of the anatomy descriptor and protocol processing.

\begin{algorithm2e}[t]
\small
\caption{ProFed training}
\label{alg:profed}
\DontPrintSemicolon
\KwIn{client data $\{\mathcal D_k\}$; rounds $T$; local epochs $E$}
Cache anatomy descriptors from full-dose training images\;
Initialize shared reconstruction/protocol weights and local anatomy heads/gates\;
\For{$t=1,\ldots,T$}{
  Server broadcasts shared weights\;
  \ForEach{client $k$ in parallel}{
    Retain local $H_\alpha$ and $\mathrm{FC}_g$\;
    \For{$e=1,\ldots,E$}{
      Reconstruct batches using cached descriptors and protocol metadata\;
      Evaluate image and projection losses\;
      Update local model using Eq.~\eqref{eq:objective}\;
    }
    Estimate $u_k$ using validation MC-dropout passes\;
    Send $(\theta_k^{\mathrm{sh}},u_k,N_k)$ to server\;
  }
  Aggregate shared weights using Eq.~\eqref{eq:aggregation}\;
}
\end{algorithm2e}

\section{Experiments}
\label{sec:experiments}
\subsection{Dataset, clients, and implementation}
We follow the experimental setting of SCAN-PhysFed \citep{yang2025patient} on the NIH-AAPM-Mayo Clinic 2016 LDCT benchmark \citep{mccollough2016tu}, containing 5,936 full-dose slices from ten patients. Eight patients are distributed across eight training clients without patient overlap between clients. The remaining two patients are reserved for testing and simulated under the known client protocols. Thus the main comparison evaluates held-out anatomy under acquisition settings represented in training, rather than reconstruction of the training patients or transfer to unseen protocols.

\paragraph{Acquisition simulation.}
Table~\ref{tab:protocols} reproduces the eight known protocol configurations from SCAN-PhysFed's published experimental setup. The simulation introduces Poisson photon noise and additive Gaussian electronic noise into projection measurements, with electronic-noise variance set to 10 as in that benchmark. The protocol vector uses the seven quantities in the table. These settings span both reduced photon counts and sparse angular sampling. The shared setup supports direct comparison of the reported benchmark scores; it does not reproduce all sources of variation between real institutions.

\begin{table*}[t]
\centering
\caption{Known client protocols used for training and held-out-patient evaluation, following SCAN-PhysFed \citep{yang2025patient}, Table A. NV: views; NDB: detector bins; PL: pixel length; DBL: detector-bin length; DSR/DDR: source/detector distance to rotation center; PN: incident photons. DBL, DSR, and DDR are in mm.}
\label{tab:protocols}
\small\setlength{\tabcolsep}{6pt}
\begin{tabular}{lrrrrrrrr}
\toprule
 & C1 & C2 & C3 & C4 & C5 & C6 & C7 & C8\\
\midrule
NV & 1024 & 128 & 512 & 384 & 712 & 200 & 560 & 368\\
NDB & 512 & 768 & 768 & 600 & 720 & 730 & 755 & 500\\
PL & 0.66 & 0.78 & 1.00 & 1.40 & 0.60 & 0.88 & 1.20 & 1.00\\
DBL & 0.72 & 0.58 & 1.20 & 1.50 & 0.82 & 0.78 & 1.30 & 1.30\\
DSR & 250 & 350 & 500 & 350 & 300 & 350 & 300 & 350\\
DDR & 250 & 300 & 400 & 300 & 350 & 280 & 400 & 350\\
PN ($\times10^5$) & 1.00 & 10.00 & 0.50 & 1.25 & 1.30 & 9.00 & 0.45 & 1.45\\
\bottomrule
\end{tabular}
\end{table*}

Experiments use RED-CNN \citep{chen2017low} and Uformer \citep{wang2022uformer} as reconstruction backbones. Training uses PyTorch, Adam \citep{kingma2014adam}, an initial learning rate of $10^{-3}$, batch size 20, and 200 communication rounds, matching the benchmark optimizer and communication settings. ProFed experiments use four NVIDIA H100 GPUs. The method uses paired low-/full-dose supervision and projection consistency.

\paragraph{Comparators.}
The CNN comparison contains twelve baselines: FedAvg, FedProx, FedNova, MOON \citep{li2021model}, FedDG \citep{liu2021feddg}, FedKD \citep{wu2022communication}, FedPer, FedBN, FedMRI, HyperFed, FedFDD, and SCAN-PhysFed. Ten of these have corresponding Transformer results; FedDG and FedBN are not reported in that setting. We distinguish these cross-method comparisons from the selected within-ProFed component comparisons below.

\subsection{Image-quality evaluation}
PSNR measures fidelity to the full-dose reference, and SSIM measures structural similarity \citep{mudeng2022prospects}. We report PSNR in dB and SSIM as a percentage. Tables~\ref{tab:cnn} and~\ref{tab:tf} include all available per-client protocol values on the held-out patients. Their average columns are computed as an unweighted mean over the eight displayed scores, so each protocol contributes equally. Values are rounded to two decimals after averaging. This explicit aggregation also avoids mixing differently rounded summary values.

These are descriptive comparisons: repeat-seed variability and patient-level confidence intervals are not available. Adjacent slices are correlated, so their number must not be treated as the number of independent clinical samples. PSNR and SSIM are reference-fidelity measures, not substitutes for a diagnostic reader study.

\begin{table*}[t]
\centering
\caption{RED-CNN: PSNR (dB) and SSIM (\%) on held-out patients under the eight known client protocols. Avg is the unweighted mean of the eight displayed client values, rounded to two decimals; bold indicates the best value in each column. Baseline client scores follow the published benchmark in \cite{yang2025patient}. Twelve comparison methods are available.}
\label{tab:cnn}
\scriptsize\setlength{\tabcolsep}{2.4pt}
\resizebox{\textwidth}{!}{%
\begin{tabular}{l cc cc cc cc cc cc cc cc cc}
\toprule
 & \multicolumn{2}{c}{C1} & \multicolumn{2}{c}{C2} & \multicolumn{2}{c}{C3} & \multicolumn{2}{c}{C4} & \multicolumn{2}{c}{C5} & \multicolumn{2}{c}{C6} & \multicolumn{2}{c}{C7} & \multicolumn{2}{c}{C8} & \multicolumn{2}{c}{Avg} \\
Method & PSNR & SSIM & PSNR & SSIM & PSNR & SSIM & PSNR & SSIM & PSNR & SSIM & PSNR & SSIM & PSNR & SSIM & PSNR & SSIM & PSNR & SSIM \\
\midrule
FedAvg & 29.99 & 82.20 & 32.82 & 78.23 & 34.68 & 86.99 & 32.30 & 80.63 & 36.87 & 92.06 & 37.36 & 89.02 & 32.73 & 80.66 & 35.46 & 87.65 & 34.03 & 84.68 \\
FedProx & 29.94 & 83.12 & 32.69 & 75.46 & 36.95 & 91.58 & 32.32 & 80.91 & 38.95 & 95.34 & 37.58 & 88.13 & 32.94 & 81.62 & 35.86 & 88.65 & 34.65 & 85.60 \\
FedNova & 29.95 & 82.29 & 32.57 & 80.57 & 37.77 & 92.16 & 32.09 & 78.98 & 34.42 & 88.69 & 36.05 & 89.57 & 32.81 & 81.41 & 35.89 & 88.71 & 33.94 & 85.30 \\
MOON & 32.50 & 75.61 & 35.55 & 85.41 & 35.72 & 80.30 & 35.69 & 82.10 & 35.38 & 78.47 & 36.31 & 84.35 & 35.70 & 81.17 & 35.62 & 80.40 & 35.31 & 80.98 \\
FedDG & 33.37 & 81.40 & 34.01 & 82.39 & 33.56 & 76.37 & 34.43 & 81.49 & 34.06 & 76.70 & 33.97 & 80.17 & 34.75 & 80.87 & 34.25 & 79.23 & 34.05 & 79.83 \\
FedKD & 29.74 & 83.11 & 32.17 & 71.65 & 36.65 & 90.93 & 32.10 & 79.19 & 40.58 & 96.71 & 37.07 & 82.55 & 32.88 & 81.30 & 36.28 & 89.08 & 34.68 & 84.32 \\
FedPer & 33.23 & 88.20 & 35.06 & 84.89 & 38.79 & 92.70 & 35.58 & 86.12 & 41.76 & 96.97 & 36.85 & 89.96 & 35.69 & 86.30 & 38.48 & 91.69 & 36.93 & 89.60 \\
FedBN & 32.55 & 84.59 & 34.76 & 84.15 & 38.18 & 90.30 & 34.77 & 83.30 & 41.24 & 96.11 & 38.66 & 91.19 & 34.74 & 83.35 & 37.89 & 90.35 & 36.60 & 87.92 \\
FedMRI & 33.28 & 83.43 & 34.51 & 85.81 & 38.62 & 92.47 & 35.86 & 87.21 & 41.49 & 96.27 & 38.22 & 91.32 & 35.87 & 86.34 & 38.09 & 91.94 & 36.99 & 89.35 \\
HyperFed & 33.74 & 88.57 & 34.83 & 85.05 & 38.87 & 92.78 & 35.44 & 85.40 & 42.94 & 97.48 & 39.15 & 92.96 & 34.99 & 85.19 & 38.12 & 91.29 & 37.26 & 89.84 \\
FedFDD & 36.18 & 93.67 & 38.43 & 94.19 & 39.72 & 96.47 & 38.51 & 95.30 & 45.56 & 98.73 & 42.72 & 97.61 & 40.74 & 96.01 & 38.54 & 94.39 & 40.05 & 95.80 \\
SCAN-PhysFed & 34.49 & 96.40 & 39.21 & 96.09 & 40.45 & 97.93 & 40.51 & 97.61 & 47.81 & \textbf{99.14} & \textbf{44.08} & 98.51 & 42.54 & 97.86 & 40.02 & 97.37 & 41.14 & 97.61 \\
\midrule
ProFed (ours) & \textbf{36.63} & \textbf{97.18} & \textbf{41.79} & \textbf{97.88} & \textbf{42.43} & \textbf{98.31} & \textbf{42.29} & \textbf{98.25} & \textbf{47.87} & 98.91 & 43.82 & \textbf{98.54} & \textbf{42.94} & \textbf{98.36} & \textbf{42.69} & \textbf{98.19} & \textbf{42.56} & \textbf{98.20} \\
\bottomrule
\end{tabular}}
\end{table*}

\subsection{Quantitative comparisons}
With RED-CNN, ProFed reaches 42.56~dB PSNR and 98.20\% SSIM averaged across clients. SCAN-PhysFed, the strongest comparator by average PSNR, reaches 41.14~dB and 97.61\%. The PSNR difference is 1.42~dB. With Uformer, ProFed achieves 44.83~dB and 98.61\%, compared with 43.65~dB and 97.02\% for SCAN-PhysFed, a 1.18~dB improvement. Gains with two backbones suggest that the combination is useful beyond a single reconstruction architecture within this benchmark.

The client-level results qualify the averages. Against SCAN-PhysFed, CNN PSNR improves on seven of eight clients, but drops from 44.08 to 43.82~dB on C6. CNN SSIM is also lower on C5 despite a small PSNR increase. With Uformer, PSNR improves on six of eight clients; SCAN-PhysFed is stronger on C3 and C5. Thus the method does not uniformly dominate either across clients or across quality metrics. The aggregate improvement should be read together with this variation, rather than as a guarantee for every protocol.

The differences between FedAvg, specialized reconstruction methods, and ProFed reflect complete method configurations. They do not isolate a causal contribution of projection supervision by themselves. In particular, the benchmark FedAvg score is not used as the reference point for the within-model ablation.

\begin{table*}[t]
\centering
\caption{Uformer: PSNR (dB) and SSIM (\%) on held-out patients under the eight known client protocols. Avg is the unweighted mean of the eight displayed client values, rounded to two decimals; bold indicates the best value in each column. Baseline client scores follow the published benchmark in \cite{yang2025patient}. Ten comparison methods are available; FedDG and FedBN were not reported for this backbone.}
\label{tab:tf}
\scriptsize\setlength{\tabcolsep}{2.4pt}
\resizebox{\textwidth}{!}{%
\begin{tabular}{l cc cc cc cc cc cc cc cc cc}
\toprule
 & \multicolumn{2}{c}{C1} & \multicolumn{2}{c}{C2} & \multicolumn{2}{c}{C3} & \multicolumn{2}{c}{C4} & \multicolumn{2}{c}{C5} & \multicolumn{2}{c}{C6} & \multicolumn{2}{c}{C7} & \multicolumn{2}{c}{C8} & \multicolumn{2}{c}{Avg} \\
Method & PSNR & SSIM & PSNR & SSIM & PSNR & SSIM & PSNR & SSIM & PSNR & SSIM & PSNR & SSIM & PSNR & SSIM & PSNR & SSIM & PSNR & SSIM \\
\midrule
FedAvg & 34.31 & 87.88 & 38.06 & 90.54 & 43.51 & 97.56 & 41.05 & 95.43 & 45.24 & 98.47 & 43.15 & 97.09 & 41.45 & 95.95 & 42.88 & 96.16 & 41.21 & 94.89 \\
FedProx & 34.32 & 87.01 & 37.74 & 88.89 & 43.73 & 97.95 & 41.12 & 93.92 & 45.39 & 98.61 & 42.99 & 97.08 & 41.38 & 94.31 & 42.10 & 94.55 & 41.10 & 94.04 \\
FedNova & 29.55 & 78.44 & 31.38 & 68.13 & 37.07 & 82.47 & 31.79 & 73.45 & 40.61 & 88.80 & 35.46 & 79.77 & 32.38 & 73.66 & 35.77 & 82.22 & 34.25 & 78.37 \\
MOON & 34.28 & 93.20 & 38.09 & 91.17 & 43.64 & 97.77 & 41.09 & 95.66 & 45.25 & 98.43 & 42.82 & 96.91 & 41.42 & 95.97 & 42.20 & 96.18 & 41.10 & 95.66 \\
FedKD & 35.92 & 89.93 & 38.70 & 91.21 & 44.26 & 98.13 & 42.30 & 95.99 & 45.93 & 98.75 & 43.57 & 97.41 & 42.54 & 96.22 & 43.51 & 96.23 & 42.09 & 95.48 \\
FedPer & 35.37 & 93.18 & 39.03 & 93.55 & 42.75 & 97.13 & 40.88 & 96.20 & 46.55 & 98.83 & 42.87 & 96.68 & 41.87 & 96.70 & 40.79 & 94.78 & 41.26 & 95.88 \\
FedMRI & 34.60 & 93.67 & 38.04 & 94.55 & 42.94 & 97.44 & 39.34 & 95.78 & 46.79 & 98.91 & 42.82 & 97.30 & 40.81 & 95.32 & 40.85 & 95.70 & 40.77 & 96.08 \\
HyperFed & 34.60 & 87.59 & 38.62 & 91.08 & 44.13 & 97.61 & 42.25 & 96.24 & 45.92 & 98.59 & 43.71 & 97.37 & 42.44 & 96.66 & 43.56 & 96.63 & 41.90 & 95.22 \\
FedFDD & 37.35 & \textbf{96.20} & 39.67 & 95.18 & 44.31 & 98.26 & 39.00 & 95.21 & 47.89 & 99.20 & 43.81 & 98.06 & 42.17 & 97.40 & 40.50 & 95.49 & 41.84 & 96.88 \\
SCAN-PhysFed & 38.55 & 94.06 & 41.74 & 96.20 & \textbf{45.52} & 98.52 & 42.48 & 96.55 & \textbf{49.50} & 99.39 & 45.32 & 97.93 & 43.35 & 96.73 & 42.71 & 96.79 & 43.65 & 97.02 \\
\midrule
ProFed (ours) & \textbf{39.26} & 96.18 & \textbf{44.33} & \textbf{98.63} & 44.90 & \textbf{98.95} & \textbf{44.60} & \textbf{98.89} & 48.91 & \textbf{99.40} & \textbf{46.58} & \textbf{99.02} & \textbf{45.30} & \textbf{98.97} & \textbf{44.76} & \textbf{98.84} & \textbf{44.83} & \textbf{98.61} \\
\bottomrule
\end{tabular}}
\end{table*}

\subsection{Evaluation on four unseen protocols}
\label{sec:unseen}
We additionally report the CNN evaluation on four acquisition configurations absent from the training federation. Following the SCAN-PhysFed setup \citep{yang2025patient}, the two held-out patients are simulated under U1--U4 in Table~\ref{tab:unseen-protocols}. This setting combines held-out anatomy with acquisition configurations not represented by C1--C8; the four client labels denote protocol domains, not four additional patients.

\begin{table}[t]
\centering
\caption{Unseen protocols from SCAN-PhysFed \citep{yang2025patient}, Table B. Abbreviations and units follow Table~\ref{tab:protocols}.}
\label{tab:unseen-protocols}
\small\setlength{\tabcolsep}{5pt}
\begin{tabular}{lrrrr}
\toprule
 & U1 & U2 & U3 & U4\\
\midrule
NV & 768 & 428 & 100 & 896\\
NDB & 550 & 590 & 768 & 730\\
PL & 0.57 & 1.10 & 0.50 & 0.70\\
DBL & 0.83 & 1.10 & 0.60 & 0.93\\
DSR & 200 & 350 & 200 & 250\\
DDR & 300 & 300 & 250 & 400\\
PN ($\times10^5$) & 1.30 & 1.40 & 11.00 & 0.90\\
\bottomrule
\end{tabular}
\end{table}

Table~\ref{tab:unseen} reports all four results and their arithmetic means. ProFed achieves 39.17~dB PSNR and 93.33\% SSIM, compared with 38.55~dB and 93.04\% for SCAN-PhysFed. Its PSNR gains over that comparator are 0.49, 0.54, 0.89, and 0.56~dB on U1--U4, respectively, averaging 0.62~dB. The average advantage is smaller than the 1.42~dB observed under the known CNN protocols. ProFed has the highest average for both metrics, but is not best in every column: HyperFed has higher PSNR on U4, and MOON has higher SSIM on U3.

The published benchmark evaluates generic FL methods using their shared model, averages local-model performance for the personalized baselines, and uses protocol vector quantization for SCAN-PhysFed \citep{yang2025patient}. 

\begin{table*}[t]
\centering
\caption{RED-CNN evaluation under four unseen protocols on the two held-out patients. PSNR (dB) and SSIM (\%). Individual values are retained from the original evaluation; Avg is recomputed as the unweighted mean across U1--U4. Baseline scores follow \cite{yang2025patient}; bold indicates column maxima.}
\label{tab:unseen}
\scriptsize\setlength{\tabcolsep}{2.4pt}
\resizebox{0.86\textwidth}{!}{%
\begin{tabular}{l cc cc cc cc cc}
\toprule
 & \multicolumn{2}{c}{U1} & \multicolumn{2}{c}{U2} & \multicolumn{2}{c}{U3} & \multicolumn{2}{c}{U4} & \multicolumn{2}{c}{Avg} \\
Method & PSNR & SSIM & PSNR & SSIM & PSNR & SSIM & PSNR & SSIM & PSNR & SSIM \\
\midrule
FedAvg & 35.57 & 88.68 & 34.00 & 85.26 & 32.52 & 76.71 & 36.86 & 91.44 & 34.74 & 85.52 \\
FedProx & 36.69 & 91.16 & 34.30 & 86.39 & 32.58 & 72.13 & 39.13 & 94.87 & 35.68 & 86.14 \\
FedNova & 35.45 & 88.83 & 34.23 & 86.29 & 32.39 & 69.20 & 35.40 & 89.99 & 34.37 & 83.58 \\
MOON & 35.21 & 78.19 & 35.30 & 79.81 & 34.55 & \textbf{83.07} & 35.56 & 78.94 & 35.16 & 80.00 \\
FedDG & 34.86 & 79.54 & 34.79 & 80.91 & 33.15 & 80.31 & 34.49 & 77.42 & 34.32 & 79.55 \\
FedKD & 36.82 & 91.60 & 34.27 & 86.37 & 32.49 & 70.54 & 40.73 & 95.90 & 36.08 & 86.10 \\
FedPer & 36.86 & 88.58 & 35.46 & 87.02 & 31.96 & 71.64 & 38.56 & 90.90 & 35.71 & 84.54 \\
FedBN & 37.84 & 89.95 & 36.16 & 88.46 & 33.67 & 81.63 & 40.21 & 94.89 & 36.97 & 88.73 \\
FedMRI & 31.12 & 82.42 & 31.95 & 79.95 & 31.72 & 71.96 & 32.04 & 81.63 & 31.71 & 78.99 \\
HyperFed & 36.18 & 89.60 & 33.47 & 82.49 & 31.94 & 65.52 & \textbf{41.70} & 94.85 & 35.82 & 83.12 \\
FedFDD & 37.80 & 92.33 & 37.06 & 91.77 & 33.71 & 80.26 & 40.53 & 95.12 & 37.28 & 89.87 \\
SCAN-PhysFed & 40.77 & 97.69 & 38.77 & 95.75 & 34.38 & 81.80 & 40.27 & 96.92 & 38.55 & 93.04 \\
\midrule
ProFed (ours) & \textbf{41.26} & \textbf{97.85} & \textbf{39.31} & \textbf{96.06} & \textbf{35.27} & 82.23 & 40.83 & \textbf{97.18} & \textbf{39.17} & \textbf{93.33} \\
\bottomrule
\end{tabular}}
\end{table*}

\subsection{Selected component comparisons}
Table~\ref{tab:components} retains the comparisons in which both protocol and anatomy branches are present. Adding projection supervision to the configuration with both branches and uncertainty weighting changes PSNR from 41.16 to 42.56~dB. With projection supervision and both branches enabled, adding uncertainty weighting changes PSNR from 42.43 to 42.56~dB. These comparisons suggest that projection supervision has the larger effect in this component study, while uncertainty weighting provides a smaller increment. No statistical significance is inferred from these point estimates.

\begin{table}[t]
\centering
\caption{Selected CNN component comparisons with anatomy and protocol conditioning enabled in every row. These are within-ProFed configurations, not reimplementations of the benchmark FedAvg baseline.}
\label{tab:components}
\small
\begin{tabular}{lcc}
\toprule
Configuration & Projection & PSNR (dB)\\
\midrule
Both branches + uncertainty & -- & 41.16\\
Both branches & $\checkmark$ & 42.43\\
Both branches + uncertainty & $\checkmark$ & \textbf{42.56}\\
\bottomrule
\end{tabular}
\end{table}

The study does not establish that protocol and anatomy become statistically disentangled, nor does it separately validate the captioning pathway or the gate's response to incorrect descriptors. Those are distinct questions from whether the complete restoration system improves reference fidelity.

\subsection{Qualitative comparisons}
\begin{figure*}[t]
\centering
\includegraphics[width=\textwidth]{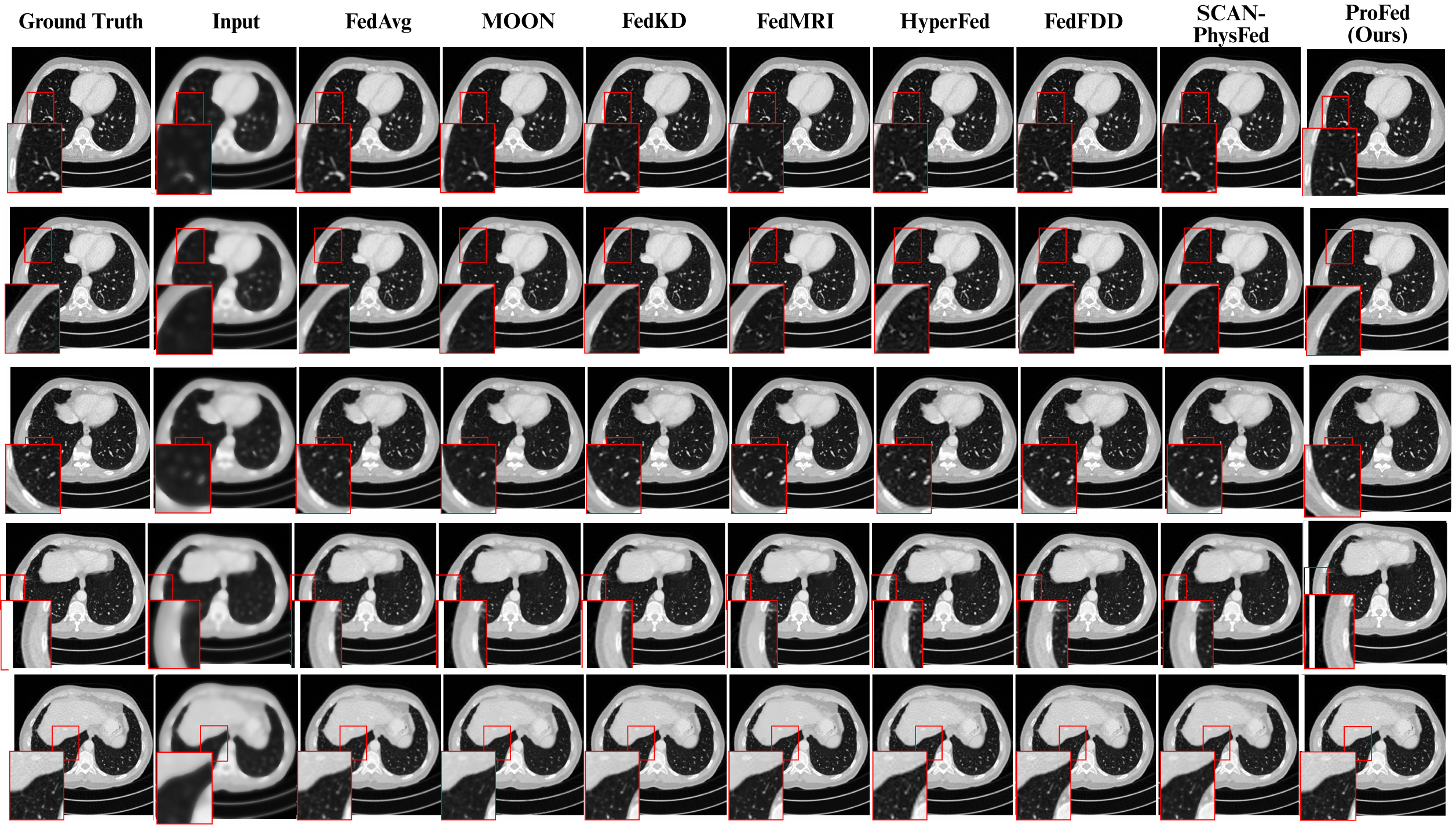}
\caption{RED-CNN qualitative comparison. Rows correspond to clients C2, C3, C5, C6, and C7; the same regions are enlarged for every method. Selected examples illustrate restoration behavior and do not establish diagnostic performance.}
\label{fig:cnn}
\end{figure*}

\begin{figure*}[t]
\centering
\includegraphics[width=\textwidth]{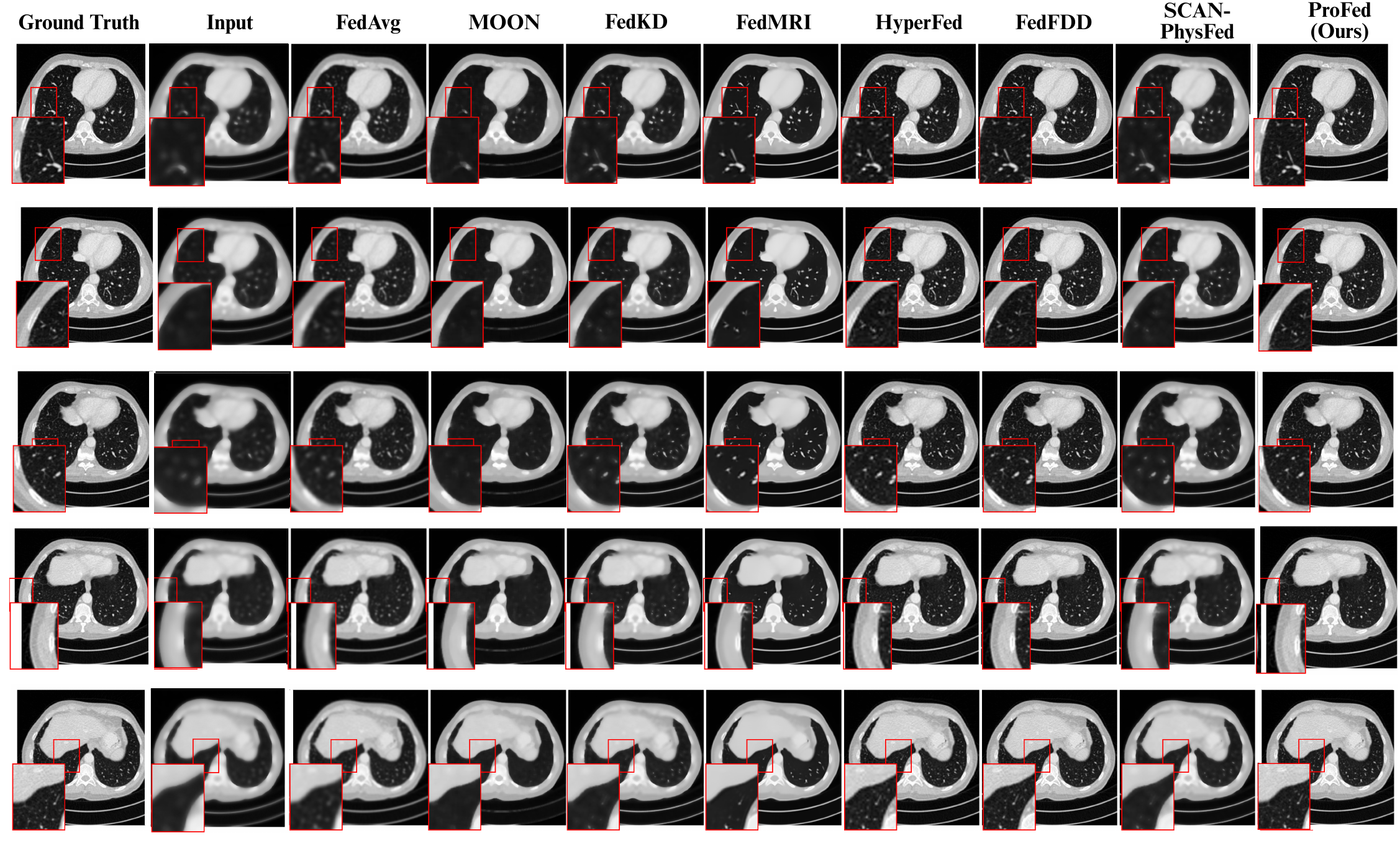}
\caption{Uformer qualitative comparison on clients C2, C3, C5, C6, and C7. ProFed and the baselines are compared against the same full-dose reference regions. Client-level numerical results, including cases where another method has higher PSNR, appear in Table~\ref{tab:tf}.}
\label{fig:tf}
\end{figure*}

Figures~\ref{fig:cnn} and~\ref{fig:tf} show five client examples for each backbone, with matching image regions across methods. ProFed produces cleaner reconstructions than several generic FL comparators and retains visible tissue boundaries in the selected regions. Differences from SCAN-PhysFed are more subtle, consistent with the smaller numerical margin over this specialized baseline.

These examples illustrate appearance rather than establish perceptual superiority. The visual comparisons do not assess whether a retained texture corresponds to real anatomy, whether a small lesion is preserved, or which method a radiologist would prefer. They complement the full-reference metrics and the explicitly reported client-level exceptions.

\section{Discussion and Limitations}
\label{sec:limitations}
\paragraph{Conditioning and deployment.}
Training uses anatomy descriptors derived from full-dose training images. Deployment requires descriptors computed exclusively from information available at inference. The reported evaluation does not separately audit test-time descriptor provenance or sensitivity to the change from clean to noisy descriptor inputs.

\paragraph{Cohort and experimental scope.}
The patient-disjoint evaluation contains only two independent test patients, each exposed to several simulated protocols. Neither the eight known protocols nor the four unseen protocols constitute independent clinical cohorts. Patient identifiers, local-epoch counts, dropout settings, captioner/checkpoint details, and repeated-seed measurements would further improve reproducibility beyond the specified benchmark settings. The unseen-client adaptation procedure also requires confirmation, as discussed in Section~\ref{sec:unseen}. Broader real-acquisition cohorts are needed to estimate performance across scanner vendors, reconstruction kernels, and institutions.

\paragraph{Physics, cost, and privacy.}
Local projection supervision requires acquisition data and a suitable forward model. Discrete operators and simulated noise do not capture every scanner effect. End-to-end inference must account for conditioning costs as well as reconstruction. Images and anatomy heads remain local, but shared updates can contain information about training data; no formal differential-privacy guarantee is provided.

\paragraph{Clinical image quality.}
Before clinical use, evaluation should examine lesion conspicuity, attenuation accuracy, resolution/noise trade-offs, and reader preference under real acquisition shifts. Potential benefits from lower-dose restoration must be balanced against the risk of suppressing subtle findings or introducing misleading texture. The present study makes no claim of diagnostic equivalence to full-dose CT.

\section{Conclusion}
ProFed combines personalized image-domain reconstruction with client-local projection supervision and uncertainty-weighted federated aggregation. Following the SCAN-PhysFed benchmark's patient-disjoint split, it improves average PSNR under eight known protocols with both RED-CNN and Uformer, while the per-protocol results expose cases where the comparator remains stronger. The four-unseen-protocol evaluation reports a smaller average CNN advantage of 0.62~dB over SCAN-PhysFed, subject to the adaptation-protocol limitation in Section~\ref{sec:unseen}. The selected component comparisons support further investigation of projection supervision as a useful element of federated restoration. A fully specified test-time conditioning pipeline and larger real-acquisition studies are necessary next steps toward deployment.

\paragraph*{Data availability.}
The study uses the Mayo Clinic LDCT benchmark \citep{mccollough2016tu}, subject to its distributor's access conditions. The reported client-level measurements, method equations, and selected component comparisons are included in this paper. 

\FloatBarrier
{\small\bibliographystyle{ieeenat_fullname}\bibliography{references}}

@inproceedings{li2019super,
  title={SUPER learning: a supervised-unsupervised framework for low-dose CT image reconstruction},
  author={Li, Zhipeng and Ye, Siqi and Long, Yong and Ravishankar, Saiprasad},
  booktitle={Proceedings of the IEEE/CVF International Conference on Computer Vision Workshops},
  year={2019}
}

@inproceedings{yang2025patient,
  title={Patient-level anatomy meets scanning-level physics: Personalized federated low-dose ct denoising empowered by large language model},
  author={Yang, Ziyuan and Chen, Yingyu and Wang, Zhiwen and Shan, Hongming and Chen, Yang and Zhang, Yi},
  booktitle={Proceedings of the Computer Vision and Pattern Recognition Conference},
  pages={5154--5163},
  year={2025}
}

@inproceedings{bera2023self,
  title={Self supervised low dose computed tomography image denoising using invertible network exploiting inter slice congruence},
  author={Bera, Sutanu and Biswas, Prabir Kumar},
  booktitle={Proceedings of the IEEE/CVF winter conference on applications of computer vision},
  pages={5614--5623},
  year={2023}
}

@article{gao2023corediff,
  title={CoreDiff: Contextual error-modulated generalized diffusion model for low-dose CT denoising and generalization},
  author={Gao, Qi and Li, Zilong and Zhang, Junping and Zhang, Yi and Shan, Hongming},
  journal={IEEE Transactions on Medical Imaging},
  volume={43},
  number={2},
  pages={745--759},
  year={2023},
  publisher={IEEE}
}

@inproceedings{zhang2021transct,
  title={TransCT: dual-path transformer for low dose computed tomography},
  author={Zhang, Zhicheng and Yu, Lequan and Liang, Xiaokun and Zhao, Wei and Xing, Lei},
  booktitle={International Conference on Medical Image Computing and Computer-Assisted Intervention},
  pages={55--64},
  year={2021},
  organization={Springer}
}

@article{yang2018low,
  title={Low-dose CT image denoising using a generative adversarial network with Wasserstein distance and perceptual loss},
  author={Yang, Qingsong and Yan, Pingkun and Zhang, Yanbo and Yu, Hengyong and Shi, Yongyi and Mou, Xuanqin and Kalra, Mannudeep K and Zhang, Yi and Sun, Ling and Wang, Ge},
  journal={IEEE transactions on medical imaging},
  volume={37},
  number={6},
  pages={1348--1357},
  year={2018},
  publisher={IEEE}
}

@article{chen2017low,
  title={Low-dose CT with a residual encoder-decoder convolutional neural network},
  author={Chen, Hu and Zhang, Yi and Kalra, Mannudeep K and Lin, Feng and Chen, Yang and Liao, Peixi and Zhou, Jiliu and Wang, Ge},
  journal={IEEE transactions on medical imaging},
  volume={36},
  number={12},
  pages={2524--2535},
  year={2017},
  publisher={IEEE}
}

@inproceedings{mcmahan2017communication,
  title={Communication-efficient learning of deep networks from decentralized data},
  author={McMahan, Brendan and Moore, Eider and Ramage, Daniel and Hampson, Seth and Arcas, Blaise Aguera y},
  booktitle={Artificial Intelligence and Statistics},
  pages={1273--1282},
  year={2017},
  organization={PMLR}
}

@inproceedings{li2020federated,
  title={Federated optimization in heterogeneous networks},
  author={Li, Tian and Sahu, Anit Kumar and Zaheer, Manzil and Sanjabi, Maziar and Talwalkar, Ameet and Smith, Virginia},
  booktitle={Proc. Mach. Learn. Syst. (MLSys)},
  volume={2},
  pages={429--450},
  year={2020}
}

@article{wang2020tackling,
  title={Tackling the objective inconsistency problem in heterogeneous federated optimization},
  author={Wang, Jianyu and Liu, Qinghua and Liang, Hao and Joshi, Gauri and Poor, H Vincent},
  journal={Advances in neural information processing systems},
  volume={33},
  pages={7611--7623},
  year={2020}
}

@inproceedings{li2021model,
  title={Model-contrastive federated learning},
  author={Li, Qinbin and He, Bingsheng and Song, Dawn},
  booktitle={Proc. IEEE/CVF Conf. Comput. Vis. Pattern Recognit. (CVPR)},
  pages={10713--10722},
  year={2021}
}

@inproceedings{liu2021feddg,
  title={Feddg: Federated domain generalization on medical image segmentation via episodic learning in continuous frequency space},
  author={Liu, Quande and Chen, Cheng and Qin, Jing and Dou, Qi and Heng, Pheng-Ann},
  booktitle={Proceedings of the IEEE/CVF conference on computer vision and pattern recognition},
  pages={1013--1023},
  year={2021}
}

@article{wu2022communication,
  title={Communication-efficient federated learning via knowledge distillation},
  author={Wu, Chuhan and Wu, Fangzhao and Lyu, Lingjuan and Huang, Yongfeng and Xie, Xing},
  journal={Nature Communications},
  volume={13},
  number={1},
  pages={2032},
  year={2022},
  publisher={Nature Publishing Group UK London}
}

@article{arivazhagan2019federated,
  title={Federated learning with personalization layers},
  author={Arivazhagan, Manoj Ghuhan and Aggarwal, Vinay and Singh, Aaditya Kumar and others},
  journal={arXiv preprint arXiv:1912.00818},
  year={2019}
}

@inproceedings{li2021fedbn,
  title={FedBN: Federated Learning on Non-IID Features via Local Batch Normalization},
  author={Li, Xiaoxiao and JIANG, Meirui and Zhang, Xiaofei and Kamp, Michael and Dou, Qi},
  booktitle={International Conference on Learning Representations},
  year = {2021}
}

@article{feng2022specificity,
  title={Specificity-preserving federated learning for MR image reconstruction},
  author={Feng, Chun-Mei and Yan, Yunlu and Wang, Shanshan and Xu, Yong and Shao, Ling and Fu, Huazhu},
  journal={IEEE Transactions on Medical Imaging},
  volume={42},
  number={7},
  pages={2010--2021},
  year={2022},
  publisher={IEEE}
}

@article{yang2023hypernetwork,
  title={Hypernetwork-based physics-driven personalized federated learning for CT imaging},
  author={Yang, Ziyuan and Xia, Wenjun and Lu, Zexin and Chen, Yingyu and Li, Xiaoxiao and Zhang, Yi},
  journal={IEEE Transactions on Neural Networks and Learning Systems},
  year={2023},
  publisher={IEEE}
}

@inproceedings{chen2024fedfdd,
  title={{FedFDD}: Federated Learning with Frequency Domain Decomposition for Low-Dose {CT} Denoising},
  author={Chen, Xuhang and Li, Zeju and Xu, Zikun and Xu, Kaijie and Ouyang, Cheng and Qin, Chen},
  booktitle={Proceedings of the International Conference on Medical Imaging with Deep Learning},
  pages={234--249}, volume={250}, series={Proceedings of Machine Learning Research}, year={2024}
}

@inproceedings{lin2019dudonet,
  title={DuDoNet: Dual domain network for CT metal artifact reduction},
  author={Lin, Wei-An and Liao, Haofu and Peng, Cheng and Sun, Xiaohang and Zhang, Jingdan and Luo, Jiebo and Chellappa, Rama and Zhou, Shaohua Kevin},
  booktitle={Proceedings of the IEEE/CVF Conference on Computer Vision and Pattern Recognition},
  pages={10512--10521},
  year={2019}
}

@article{darzidehkalani2022federated,
  title={Federated learning in medical imaging: part I: toward multicentral health care ecosystems},
  author={Darzidehkalani, Erfan and Ghasemi-Rad, Mohammad and Van Ooijen, PMA},
  journal={Journal of the american college of radiology},
  volume={19},
  number={8},
  pages={969--974},
  year={2022},
  publisher={Elsevier}
}

@article{lee2020biobert,
  title={BioBERT: a pre-trained biomedical language representation model for biomedical text mining},
  author={Lee, Jinhyuk and Yoon, Wonjin and Kim, Sungdong and Kim, Donghyeon and Kim, Sunkyu and So, Chan Ho and Kang, Jaewoo},
  journal={Bioinformatics},
  volume={36},
  number={4},
  pages={1234--1240},
  year={2020},
  publisher={Oxford University Press}
}

@article{kingma2014adam,
  title={Adam: A method for stochastic optimization},
  author={Kingma, Diederik P},
  journal={arXiv preprint arXiv:1412.6980},
  year={2014}
}

@inproceedings{wang2022uformer,
  title={Uformer: A general u-shaped transformer for image restoration},
  author={Wang, Zhendong and Cun, Xiaodong and Bao, Jianmin and Zhou, Wengang and Liu, Jianzhuang and Li, Houqiang},
  booktitle={Proceedings of the IEEE/CVF conference on computer vision and pattern recognition},
  pages={17683--17693},
  year={2022}
}

@article{mudeng2022prospects,
  title={Prospects of structural similarity index for medical image analysis},
  author={Mudeng, Vicky and Kim, Minseok and Choe, Se-woon},
  journal={Applied Sciences},
  volume={12},
  number={8},
  pages={3754},
  year={2022},
  publisher={MDPI}
}

@article{mccollough2016tu,
  title={TU-FG-207A-04: overview of the low dose CT grand challenge},
  author={McCollough, Cynthia},
  journal={Medical physics},
  volume={43},
  number={6Part35},
  pages={3759--3760},
  year={2016},
  publisher={Wiley Online Library}
}
\end{document}